# The GAMMA-400 gamma-ray telescope characteristics. Angular resolution and electrons/protons separation

**A.A. Leonov and O. Adriani, on behalf of the GAMMA-400 collaboration**
National Research Nuclear University MEPhI, Moscow
E-mail: leon@ibrae.ac.ru
University of Florence and INFN Sezione di Firenze
E-mail: adriani@fi.infn.it

**A.M. Galper[a;b], V. Bonvicini[c], N.P. Topchiev[a], O. Adriani[d], R.L. Aptekar[e],
I.V. Arkhangelskaja[b], A.I. Arkhangelskiy[b], L. Bergstrom[f], E. Berti[d], G. Bigongiari[g],
S.G. Bobkov[h], M. Boezio[c], E.A. Bogomolov[e], S. Bonechi[g], M. Bongi[d], S. Bottai[d],
G. Castellini[j], P.W. Cattaneo[k], P. Cumani[c], G.L. Dedenko[b], C. De Donato[l],
V.A. Dogiel[a], M.S. Gorbunov[h], Yu.V. Gusakov[a], B.I. Hnatyk[n], V.V. Kadilin[b], V.A. Kaplin[b],
A.A. Kaplun[b], M.D. Kheymits[b], V.E. Korepanov[o], J. Larsson[m], A.A. Leonov[b],
V.A. Loginov[b], F. Longo[c], P. Maestro[g], P.S. Marrocchesi[g], V.V. Mikhailov[b], E. Mocchiutti[c],
A.A. Moiseev[p], N. Mori[d], I.V. Moskalenko[q], P.Yu. Naumov[b], P. Papini[d], M. Pearce[m],
P.Picozza[l], A.V. Popov[h], A. Rappoldi[k], S. Ricciarini[j], M.F. Runtso[b], F. Ryde[m], O.V. Serdin[h],
R. Sparvoli[l], P. Spillantini[d], S.I. Suchkov[a], M. Tavani[r], A.A. Taraskin[b], A. Tiberio[d],
E.M. Tyurin[b], M.V. Ulanov[e], A. Vacchi[c], E. Vannuccini[d], G.I. Vasilyev[e], Yu.T. Yurkin[b],
N. Zampa[c], V.N. Zirakashvili[s] and V.G. Zverev[b]**

a Lebedev Physical Institute, Russian Academy of Sciences, Moscow, Russia
b National Research Nuclear University MEPhI, Moscow, Russia
c INFN, Sezione di Trieste and Physics Department of University of Trieste, Trieste, Italy
d INFN, Sezione di Firenze and Physics Department of University of Florence, Firenze, Italy
e Ioffe Institute, Russian Academy of Sciences, St. Petersburg, Russia
f Stockholm University, Department of Physics; and the Oskar Klein Centre, AlbaNova
University Center, Stockholm, Sweden
g Department of Physical Sciences, Earth and Environment, University of Siena and INFN,
Sezione di Pisa, Italy
h Scientific Research Institute for System Analysis, Russian Academy of Sciences, Moscow, Russia
i Research Institute for Electromechanics, Istra, Moscow region, Russia
j IFAC- CNR and Istituto Nazionale di Fisica Nucleare, Sezione di Firenze, Firenze, Italy
k INFN, Sezione di Pavia, Pavia, Italy
l INFN, Sezione di Roma 2 and Physics Department of University of Rome Tor Vergata, Italy
m KTH Royal Institute of Technology, Department of Physics; and the Oskar Klein Centre,
AlbaNova University Center, Stockholm, Sweden
n Taras Shevchenko National University of Kyiv, Kyiv, Ukraine
o Lviv Center of Institute of Space Research, Lviv, Ukraine
p CRESST/GSFC and University of Maryland, College Park, Maryland, USA
q Hansen Experimental Physics Laboratory and Kavli Institute for Particle Astrophysics and
Cosmology, Stanford University, Stanford, USA

r Istituto Nazionale di Astrofisica IASF and Physics Department of University of Rome Tor Vergata, Rome, Italy
s Pushkov Institute of Terrestrial Magnetism, Ionosphere, and Radiowave Propagation, Troitsk, Moscow region, Russia

The measurements of gamma-ray fluxes and cosmic-ray electrons and positrons in the energy range from 100 MeV to several TeV, which will be realized by the specially designed GAMMA-400 gamma-ray telescope, concern with the following broad range of scientific topics. Search for signatures of dark matter, surveying the celestial sphere in order to study point and extended sources of gamma-rays, measuring the energy spectra of Galactic and extragalactic diffuse gamma–ray emission, study of gamma-ray bursts and gamma-ray emission from the Sun, as well as high precision measurements of spectra of high-energy electrons and positrons, protons and nuclei up to the knee. To clarify these scientific problems with the new experimental data the GAMMA-400 gamma-ray telescope possesses unique physical characteristics comparing with previous and present experiments. For gamma-ray energies more than 100 GeV GAMMA-400 provides the energy resolution ~1% and angular resolution better than 0.02 deg. The methods, developed to reconstruct the direction of incident gamma photon, are presented in this paper, as well as, the capability of the GAMMA-400 gamma-ray telescope to distinguish electrons and positrons from protons in cosmic rays is investigated.

The first point concerns with the space topology of high-energy gamma photon interaction in the matter of GAMMA-400. Multiple secondary particles, generated inside gamma-ray telescope, produce significant problems to restore the direction of initial gamma photon. Also back-splash particles, i.e., charged particles and gamma photons generated in calorimeter and moved upward, mask the initial tracks of electron/positron pair from conversion of incident gamma photon. The processed methods allow us to reconstruct the direction of electromagnetic shower axis and extract the electron/positron trace. As a result, the direction of incident gamma photon with the energy of 100 GeV is calculated with an accuracy of better than 0.02 deg.

The main components of cosmic rays are protons and helium nuclei, whereas the part of lepton component in the total flux is $\sim 10^{-3}$ for high energies. The separate contribution in proton rejection is studied for each detector system of the GAMMA-400 gamma-ray telescope. Using combined information from all detector systems allow us to provide the rejection from protons with a factor of $\sim 4 \times 10^5$ for vertical incident particles and $\sim 3 \times 10^5$ for particle with initial inclination of 30 deg.

*The GAMMA-400 gamma-ray telescope characteristics. Angular resolution and electrons/protons separation*  
Alexey Leonov and Oscar Adriani

## 1. Introduction

The GAMMA-400 gamma-ray telescope is intended to measure the fluxes of gamma-rays and cosmic-ray electrons and positrons in the energy range from 100 MeV to several TeV [1]. Such measurements concern with the following scientific tasks: search for signatures of dark matter, investigation of point sources of gamma-rays, studies of the energy spectra of Galactic and extragalactic diffuse emission, studies of gamma-ray bursts and gamma-ray emission from the Sun, as well as high precision measurements of spectra of high-energy electrons and positrons, protons and nuclei up to the knee. To fulfill these measurements the GAMMA-400 gamma-ray telescope possesses unique physical characteristics in comparison with previous and present experiments.

In the present paper the methods providing the GAMMA-400 angular resolution better than 0.02 deg for gamma-ray energies more than 100 GeV are described. Moreover the capability of the GAMMA - 400 gamma ray telescope to distinguish electrons and positrons from protons in cosmic rays is investigated. Using combined information from all detector systems of gamma-ray telescope allow us to provide rejection from protons with a factor of $\approx 4 \times 10^5$ for vertical incident particles and $\approx 3 \times 10^5$ for particle with initial inclination of 30 deg.

## 2. The GAMMA-400 gamma-ray telescope

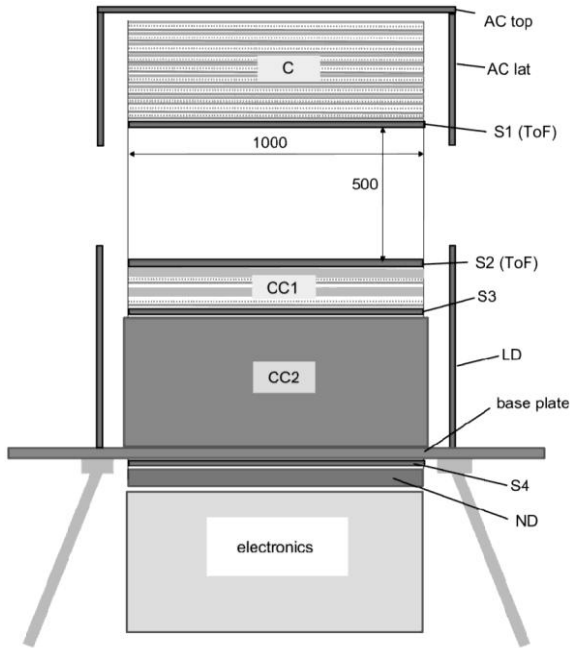

Figure 1. The GAMMA-400 physical scheme.

The GAMMA-400 physical scheme is shown in Fig. 1. GAMMA-400 consists of scintillation anticoincidence top and lateral detectors (AC), converter-tracker (C) with 10 layers of double (x, y) silicon strip coordinate detectors (pitch of 0.1 mm) interleaved with tungsten conversion foils, scintillation detectors (S1 and S2) of time-of-flight system (TOF), calorimeter from two parts (CC1 and CC2), lateral detectors LD, scintillation detectors S3 and S4 and neutron detector ND to separate hadron and electromagnetic showers. The anticoincidence detectors surrounding the converter-tracker are served to distinguish gamma-ray events from the much more numerous charged-particle events. Converter-tracker information is applied to precisely determine the direction of each incident particle and calorimeter measurements are used to determine its energy. All scintillation detectors consist from two independent layers. Each layer has width of 1 cm. The time-of-flight system, where detectors S1 and S2 are separated by approximately 500 mm, determines the top-down direction of arriving particle. Additional scintillation detectors S3 and S4 improve hadron and electromagnetic showers separation.

The imaging calorimeter CC1 consists of 2 layers of double (x, y) silicon strip coordinate detectors (pitch of 0.1 mm) interleaved with planes from CsI(Tl) crystals, and the electromagnetic calorimeter CC2 consists of CsI(Tl) crystals with the dimensions of 36 mm × 36 mm × 430 mm. The total converter-tracker thickness is about 1 $X_0$ ($X_0$ is the radiation length). The thickness of CC1 and CC2 is 3 $X_0$ and 22 $X_0$, respectively. The total calorimeter thickness is 25 $X_0$ or 1.2 $\lambda_0$ ($\lambda_0$ is nuclear interaction length). Using thick calorimeter allows us to extend the energy range up to several TeV and to reach the energy resolution up to approximately 1% above 100 GeV.

## 3. Reconstruction of incident direction of high-energy gamma in the GAMMA‑400 gamma-ray space telescope.

The methods to reconstruct the incident direction of the gamma were developed with the GEANT4 simulation toolkit software [2]. The reconstruction procedure applies the energy deposits in 12 silicon-strip layers. Each layer being composed of two plates of mutually perpendicular strips (Y and Z). 10 layers, 8 of which being interleaved with tungsten foils, composes a converter-tracker, whereas other 2 layers are set in spatially-sensitive calorimeter.

A direction-reconstruction technique based on strip energy release has been developed. The plane flux of gamma has simulated just above top AC plane. At first, for each silicon-strip layer with energy release the following procedure is applied. The distribution of the sum of energy releases in strips along strip positions is constructed (Fig. 2a). The horizontal line is median, which is calculated as a half sum of the extreme points for constructed distribution. The intersection point of median with piecewise continuous distribution gives the estimation of median energy location in silicon-strip layer (Fig. 2a).

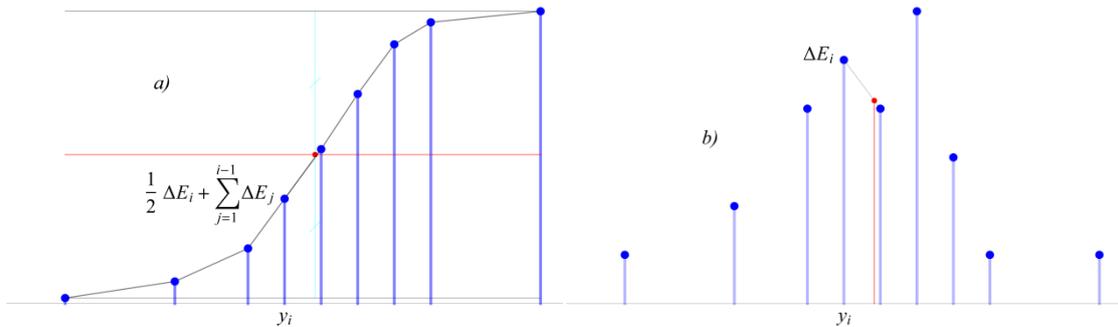

*Figure 2. The calculation of location and energy weight for median of the energy release distribution in silicon-strip layer.*

To find the energy weight of the median the ordinary distribution of energy releases in strips along strip positions is built (Fig. 2b). The median energy weight is defined using the obtained median location for the piece line linking adjacent (respective median location) points of the obtained distribution (Fig. 2b). Then the estimation of the initial direction is obtained using fitting procedure for the median locations in silicon-strip layers. Around the estimated direction the corridor from strips is constructed. The energy releases in strips outside the corridor are ignored. After that the iteration procedure starts, narrowing the corridor from strips in each silicon-strip layer, but not less, then five strip pitch. For each iteration step the median energy weight from previous iteration is taken into account.

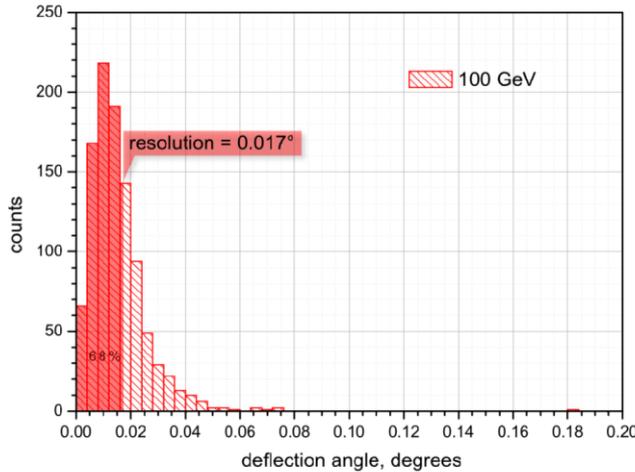

Figure 3. The distribution of space angular deflections between the direction reconstructed for each event and the distribution median value.

To calculate the angular resolution, the distribution of a space angular deflection between the direction reconstructed for each event and the median value for all events in distribution is analyzed. Such distribution for 100 GeV gamma is shown in Fig. 3. Angular resolution is defined a semiopening of the circular conical surface, containing 68% of events. The computed angular resolution taken as this value is shown in Fig.4 as function of initial energy for incident and polar angle of gamma equal $5^0$. The results are not changing significantly at least for incident angles till $15^0$. The FERMI experiment data for front section are also shown in Fig. 4 [3]. For the energies more than 10 GeV the GAMMA-400 gamma-ray telescope provides several times better angular resolution.

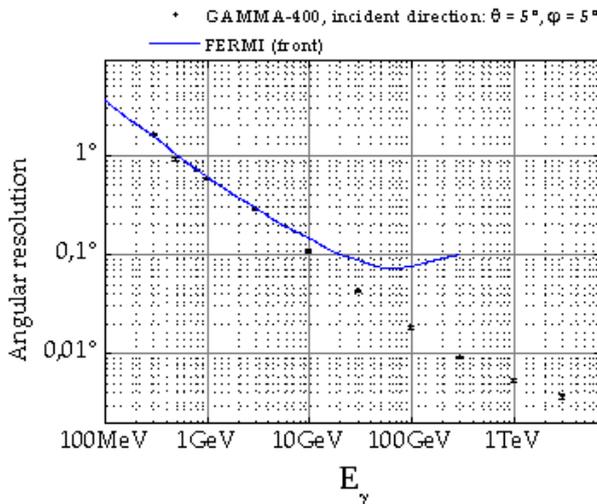

Figure 4. The energy dependence of GAMMA-400 angular resolution

### 4. Methods to reject protons from electrons using the GAMMA-400 gamma-ray telescope

Protons produce the main background, when detecting electrons in cosmic rays. The electron/proton identification usually relies on a comparison of longitudinal and transversal shower profiles and the total energy deposition in the calorimeter system on the basis of the fact that electromagnetic and hadronic showers have different spatial and energy topology view. Also the number of neutrons generated in the electromagnetic cascade is much smaller than that in the hadronic cascade. To reject protons from electrons using the GAMMA-400 instrument the information from detectors ND, S4, S3, S2, CC1, C, and CC2 is used.

The rejection factor for protons is calculated as the ratio of number of initial protons with energy more than 100 GeV, assuming that the proton energy spectrum power is -2.7, to the number of events identified as electrons with energy 100±2 GeV. Firstly, the rejection factor for vertical incident protons is evaluated. All processed criteria to suppress protons are based on selecting cutoffs to distinguish proton and electron events. The location of the cutoff for each criterion is selected in order to retain 98% of electrons. Totally 25 cutoffs are used to reject protons. With presented selection also ~30% of electrons are lost due to proton rejection.

The main contribution in the total rejection factor for protons in the GAMMA-400 telescope concerns with significantly different number of neutrons generated in the electromagnetic and hadron cascades. In cascades induced by protons, the generation of neutrons is more intensive than in the electromagnetic shower. The source of neutrons in cascades induced by electron concerns with generation of gamma rays with energy about 17 MeV. Those gamma rays, by-turn, could generate neutrons in Giant resonance reaction. Analyzing information from the neutron detector placed just under the CC2 calorimeter, it is possible to suppress protons by a factor of 400. The distributions for number of neutrons at the entrance of ND from initial electrons and protons are shown in Fig. 5 (left part). The neutron number cutoff to separate protons is equal 60.

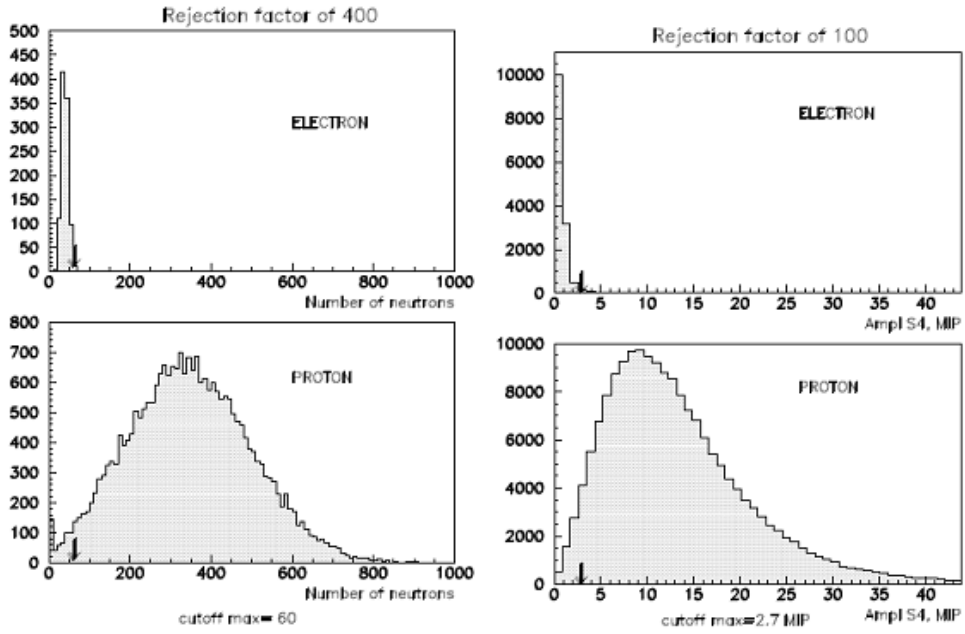

*Figure 5. The distributions for number of neutrons at the entrance of ND from initial electrons (left-top) and protons (left-bottom). The distributions for signals in S4 from initial electrons (right-top) and protons (right-bottom).*

Next significant contribution in total rejection factor for protons is provided by analyzing the energy release in S4. This rejection concerns with the difference in the attenuation for hadron and electromagnetic cascades. Such criterion was used in the PAMELA experiment data analysis [4]. The distributions for signals in S4 from initial electrons and protons are shown in Fig. 5 (right part). Selecting events with signals in S4 less than 2.7 MIP (MIP is minimum ionizing particle, 1 MIP is ~ 2 MeV for S4), it is possible to suppress protons with a factor of 100.

Additional rejection is obtained when analyzing CC2 signal. Calorimeter CC2 contains CsI(Tl) crystals with transversal dimension of 36 mm × 36 mm and longitudinal dimension of 430 mm. The criterion is based on the difference of the spatial size for hadron and electromagnetic showers. Such topology difference was successfully applied with calorimeter data in the PAMELA mission to separate electrons from antiproton sample and positrons from proton sample [5]. The distributions for the ratio of signal in the crystal containing cascade axis to the value of total signal in CC2 for initial electrons and protons are compared (Fig 6). Using the distribution for the initial electrons, the values of two cutoffs are determined: 71.25% and 74.40%. For the proton distribution, only events placed between these electron cutoffs are retained. Applying this rejection provides the rejection factor of ~30.

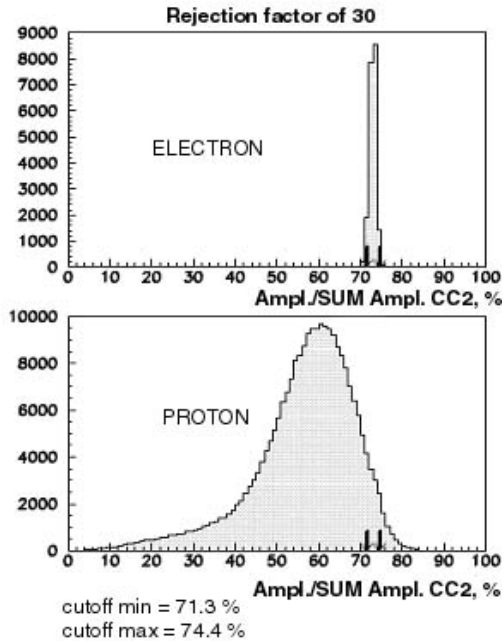

Figure 6. The distributions for the ratio of signal in the crystal containing the cascade axis to the value of total signal in CC2 for initial electrons (top) and protons (bottom).

The differences in proton and electron cascade spatial size are also used when analyzing information from silicon strips in CC1. To take into account the fact that the hadron cascade begins to develop deeper inside the instrument than the electromagnetic shower the signals in detector systems are only considered to be placed less than $4\,X_0$ inside the telescope, namely S2, S3 and CsI(Tl) crystals inCC1. All above discussed proton rejection criteria were considered separately from each other. Using all criteria in the combination, it is possible to obtain the rejection factor for protons equal to $(4\pm0.4)\times10^5$.

## 5. Conclusion

The processed methods allow us to reconstruct the direction of electromagnetic shower axis and extract the electron/positron trace. As a result, the direction of incident gamma photon with the energy of 100 GeV is calculated with an accuracy of better than 0.02 deg.

Using the combined information from all detector systems of the GAMMA-400 gamma-ray telescope, it is possible to provide effective rejection of protons from electrons. The proposed methods are based on the difference of the development of hadron and electromagnetic showers inside the instrument. It was shown that the rejection factor is several times better than $10^5$.

## References

[1] A.M. Galper, et al., *Status of the GAMMA-400 project, Adv. Space Res.,* 51, (2013), 297-300.

[2] http://geant4.cern.ch.

[3] http://www.slac.stanford.edu/exp/glast/groups/canda/lat_Performance.htm.

[4] A.V. Karelin, S.V. Borisov, S.A. Voronov, and V.V. Malakhov, *Separation of the electron and proton cosmic-ray components by means of a calorimeter in the PAMELA satellite-borne experiment for the case of particle detection within a large aperture, Physics of Atomic Nuclei, V. 76,* (2013), Issue 6, pp. 737-747.

[5] W. Menn, et al., *The PAMELA space experiment, Adv. Space Res., 51,* (2013), 209–218.